\documentstyle[12pt]{article}
\begin{document}
\title{Nearly Degenerate Mass and Bi-maximal Mixing of Neutrinos
 in the SO(3) Gauge Model of Leptons}
\author{Yue-Liang  Wu\footnote{YLWU@ITP.AC.CN}  \\  
\\
Institute of Theoretical Physics, Chinese Academy of Sciences, \\
 P.O. Box 2735, Beijing 100080, China } 
\date{AS-ITP-99-01, hep-ph/9901245, to appear in Euro. Phys. J. C.}
\maketitle

\begin{abstract} 
A realistic scheme for masses and mixing of leptons is 
investigated in the model with gauged SO(3) lepton flavor symmetry. 
Within this scheme, a nearly `bi-maximal' neutrino mixing pattern 
with maximal CP-violating phase is found to be the only possible solution 
for reconciling both solar and atmospheric neutrino flux anomalies. 
Three Majorana neutrino masses are nearly degenerate and allow to be large 
enough to play a significant cosmological role.  Lepton flavor-violating 
effects via SO(3) gauge interactions can be as large as the present 
experimental limits. Masses of the SO(3) gauge bosons are bounded to 
be above $24$ TeV when taking the SO(3) gauge boson mixing angle 
$\theta_{F}$ and coupling constant $g'_{3}$ to be the same as those 
($\theta_{W}$ and $g$) in the electroweak theory. 
\end{abstract}

\newpage

    Evidence for oscillation of atmospheric neutrinos (and hence nonzero neutrino mass) 
reported recently by the Super-Kamiokande collaboration\cite{SUPERK1} is thought as a major 
milestone in the search for new physics beyond the standard model(SM). Massive neutrinos are 
also regarded as the best candidate for hot dark matter and may play an essential role 
in the evolution of the large-scale structure of the universe\cite{HDM}. Nonzero neutrino 
mass can also provide a natural explanation on the solar neutrino problem 
which is in fact the first indication for neutrino oscillation\cite{SOLAR}. The solar neutrino 
flux measured recently by the Super-Kamiokande collaboration\cite{SUPERK2} is only about $37\%$     
of that expected from the `BP' standard solar model (`BP' SSM)\cite{BP}. The SM has been tested 
by more and more precise experiments, its greatest success is the gauge symmetry structure 
$SU_{L}(2)\times U_{Y}(1)$. Nevertheless, neutrinos are assumed to be massless in the SM.  
To introduce masses and mixings of the neutrinos, it is necessary to modify and go beyond the 
SM. As a simple extension of the SM, it is of interest to investigate possible flavor symmetries among 
three families of leptons. In the recent paper\cite{YLWU}, we have introduced gauged SO(3) 
symmetry for the three lepton families and observed that it has some remarkable features which are 
applicable to the current interesting phenomena concerning neutrinos. As the first essential step, it  
has been shown \cite{YLWU} that the SO(3) gauge symmetry allows three Majorana 
neutrino masses to be nearly degenerate\footnote{Recently, authors in refs.\cite{CM,MA,CW,BHKR} have 
also discussed SO(3) flavor symmetry in connection with nearly degenerate neutrinos.}  
and large enough for hot dark matter. 
The nearly `bi-maximal' mixing patterns (that include the bi-maximal mixing pattern\cite{BMAX,BMAXM} and 
democratic mixing pattern\cite{DEM,DEMM}) with maximal CP-violating phase were been resulted 
to reconcile both solar and atmospheric neutrino flux anomalies. As the vacuum structure 
of spontaneous SO(3) gauge symmetry breaking can automatically generate a maximal CP-violating phase, 
the scheme can be made to be consistent with the neutrinoless double beta decay and leads to 
the Georgi-Glashow form for neutrino mass matrix\cite{GG}. 
In this paper, we are going to further investigate such a gauge model and to show how 
to carry out other necessary steps to realize a realistic scheme for masses 
and mixing of the leptons. As it was expected in \cite{YLWU} that the realistic scheme 
does not significantly change the main interesting features mentioned overthere. 
In particular, it will be seen that within the realistic 
scheme presented here the `bi-maximal' mixing pattern becomes the only possible solution 
to reconcile both solar and atmospheric neutrino data.

  For a more simple and model-independent consideration, we shall start directly from the 
following SO(3)$_{F}\times$SU(2)$_{L}\times$U(1)$_{Y}$ invariant effective lagrangian for leptons 
\begin{eqnarray}
{\cal L} & = &  \frac{1}{2}g'_{3}A_{\mu}^{k}
\left( \bar{L}_{i}\gamma^{\mu} (t^{k})_{ij}L_{j} 
+ \bar{e}_{R i} \gamma^{\mu}(t^{k})_{ij}e_{R j} \right) 
+ D_{\mu}\varphi^{\ast}D^{\mu}\varphi + D_{\mu}\varphi^{'\ast}D^{\mu}\varphi' \nonumber \\
& + & \left(C_{1}\frac{\varphi_{i}\varphi_{j}}{M_{1}M_{2}} + 
C'_{1}\frac{\varphi'_{i}\varphi'_{j}}{M'_{1}M'_{2}}\frac{\chi}{M} + C''_{1} 
\frac{\chi'}{M'}\delta_{ij} \right)\bar{L}_{i} \phi_{1}e_{R\ j} + h.c.
 \\
& + & \left(C_{0}\delta_{ij} + C'_{0}\frac{\varphi_{i}
\varphi_{j}^{\ast}}{M_{2}^{2}} + C''_{0}\frac{\varphi'_{i}
\varphi_{j}^{'\ast}}{M_{2}^{'2}} \right)\frac{1}{M_{N}} 
\bar{L}_{i} \phi_{2}\phi_{2}^{T}L_{j}^{c} + h.c. + {\cal L}_{SM} \nonumber 
\end{eqnarray}
which is assumed to be resulted from integrating out heavy particles. 
Where ${\cal L}_{SM} $ denotes the lagrangian of the standard 
model. $\bar{L}_{i}(x) = (\bar{\nu}_{i}, \bar{e}_{i})_{L}$ 
(i=1,2,3) are the SU(2)$_{L}$ doublet leptons. $e_{R\ i}$ (i=1,2,3) are 
the three right-handed charged leptons. $\phi_{1}(x)$ and $\phi_{2}(x)$ 
are two Higgs doublets. $\varphi^{T}=(\varphi_{1}(x), \varphi_{2}(x), 
\varphi_{3}(x))$ and $\varphi'^{T} = (\varphi'_{1}(x), \varphi'_{2}(x), 
\varphi'_{3}(x))$ are two complex SO(3) triplet scalars. $\chi(x)$ and 
$\chi'(x)$ are two singlet scalars. $M_{1}$, $M_{2}$, $M$, 
$M'_{1}$, $M'_{2}$, $M'$ and $M_{N}$ are possible mass scales concerning 
heavy fermions. $C_{a}$, $C'_{a}$ and 
$C''_{a}$ ($a=0,1$) are six coupling constants. The structure of the above 
effective lagrangian can be obtained by imposing an additional U(1) symmetry, 
which is analogous to the construction of the $C_{0}$ and $C_{1}$ terms 
discussed in detail in ref.\cite{YLWU}. After the symmetry 
SO(3)$_{F}\times$SU(2)$_{L}\times$U(1)$_{Y}$ is broken down to the U(1)$_{em}$ 
symmetry, we obtain mass matrices of the neutrinos and charged leptons as follows
\begin{eqnarray}
(M_{e})_{ij} & = & m_{1} \frac{\hat{\sigma}_{i}\hat{\sigma}_{j}}{\sigma^{2}} 
+ m'_{1} \frac{\hat{\sigma}'_{i}\hat{\sigma}'_{j}}{\sigma^{'2}} + m''_{1} 
\delta_{ij} \nonumber \\
(M_{\nu})_{ij} & = &  m_{0} \delta_{ij} + m'_{0} \frac{\hat{\sigma}_{i}
\hat{\sigma}_{j}^{\ast} + \hat{\sigma}_{j}\hat{\sigma}_{i}^{\ast} }{2\sigma^{2}} 
+ m''_{0} \frac{\hat{\sigma}'_{i}\hat{\sigma}_{j}^{'\ast} + 
\hat{\sigma}'_{j}\hat{\sigma}_{i}^{'\ast}}{2\sigma^{'2}}  
\end{eqnarray}
where the mass matrices $M_{e}$ and $M_{\nu}$ are defined in the basis
 ${\cal L}_{M} = \bar{e}_{L}M_{e}e_{R} + 
\bar{\nu}_{L}M_{\nu}\nu^{c}_{L} + h.c. $. The constants $\hat{\sigma}_{i} = 
<\varphi_{i}(x)>$ and $\hat{\sigma}'_{i} = <\varphi'_{i}(x)>$ 
represent the vacuum expectation values of the two triplet 
scalars $\varphi(x)$ and $\varphi'(x)$. The six mass parameters are defined as:
$m_{0} = C_{0}v_{2}^{2}/M_{N}$, $m'_{0} = C'_{0}(\sigma^{2}/M_{2}^{2})(v_{2}^{2}/M_{N})$, 
$m''_{0} = C''_{0}(\sigma^{'2}/M_{2}^{'2})(v_{2}^{2}/M_{N})$,
$m_{1} = C_{1}v_{1}\sigma^{2}/M_{1}M_{2}$, $m'_{1} = C'_{1}
(\xi/M)(v_{1}\sigma^{2}/M'_{1}M'_{2})$ and $m''_{1} = C''_{1}v_{1}\xi'/M$. Here 
$\sigma = \sqrt{|\hat{\sigma}_{1}|^{2} + |\hat{\sigma}_{2}|^{2} + 
|\hat{\sigma}_{3}|^{2}}$ and $\sigma' = \sqrt{|\hat{\sigma}'_{1}|^{2} + 
|\hat{\sigma}'_{2}|^{2} + |\hat{\sigma}'_{3}|^{2}}$. 
$\xi =<\chi(x)>$ and $\xi'=<\chi'(x)>$ denote the vacuum expectation values of 
the two singlet scalars. 

  Utilizing the gauge symmetry property, it is convenient to reexpress 
the complex triplet scalar fields $\varphi_{i}(x)$ and $\varphi'_{i}(x)$ 
in terms of the SO(3) rotational fields $O(x)= e^{i\eta_{i}(x)t^{i}},\  
O'(x)= e^{i\eta'_{i}(x)t^{i}} \in $ SO(3) 
\begin{eqnarray} 
& & \left( \begin{array}{c}
  \varphi_{1}(x) \\
  \varphi_{2}(x)   \\
  \varphi_{3}(x)   \\
\end{array} \right) = e^{i\eta_{i}(x)t^{i}} \frac{1}{\sqrt{2}}
\left( \begin{array}{c}
  \rho_{1}(x) \\
  i\rho_{2}(x)   \\
  \rho_{3}(x)   \\
\end{array} \right) \nonumber \\
& & \left( \begin{array}{c}
  \varphi'_{1}(x) \\
  \varphi'_{2}(x)   \\
  \varphi'_{3}(x)   \\
\end{array} \right) = e^{i\eta'_{i}(x)t^{i}} \frac{1}{\sqrt{2}}
\left( \begin{array}{c}
  \rho'_{1}(x) \\
  i\rho'_{2}(x)   \\
  \rho'_{3}(x)   \\
\end{array} \right)    
\end{eqnarray}
where the three rotational fields $\eta_{i}(x)$ ($\eta'_{i}(x)$) and 
the three amplitude fields $\rho_{i}(x)$ ($\rho'_{i}(x)$) reparameterize 
the six real fields of the complex triplet scalar field $\varphi(x)$ 
($\varphi(x)$). Here the imaginary part is assigned to the second amplitude 
field\footnote{For other two possible assignments and correponding 
consequences will be discussed elsewhere\cite{YLWU2}} . 
 SO(3) gauge symmetry allows one to remove three degrees 
of freedom from the six rotational fields. Thus the vacuum structure of the 
SO(3) symmetry is determined only by nine degrees of freedom.
These nine degrees of freedom can be taken as $\rho_{i}(x)$, $\rho'_{i}(x)$ 
and $(\eta_{i}(x) - \eta'_{i}(x))$ without lossing generality. Here we will consider 
the following vacuum structure for the SO(3) symmetry breaking
\begin{equation}
<\rho_{i}(x)> = \sigma_{i}, \qquad <\rho'_{i}(x)> = \sigma'_{i}, \qquad 
<(\eta_{i}(x) - \eta'_{i}(x))> = 0
\end{equation}
With this vacuum structure, the mass matrices of the neutrinos and charged 
leptons can be reexpressed as 
\begin{eqnarray}
M_{e} & = & m_{1}\left( \begin{array}{ccc}
  s_{1}^{2}s_{2}^{2} & ic_{1}s_{1}s_{2}^{2} & s_{1}c_{2}s_{2}  \\
   ic_{1}s_{1}s_{2}^{2} & -c_{1}^{2}s_{2}^{2} &  ic_{1}c_{2}s_{2} \\
  s_{1}c_{2}s_{2} & ic_{1}c_{2}s_{2}
 & c_{2}^{2}s_{2}^{2}  \\ 
\end{array} \right) \nonumber \\
& + & m'_{1}\left( \begin{array}{ccc}
  s_{1}^{'2}s_{2}^{'2} & ic'_{1}s'_{1}s_{2}^{'2} & s'_{1}c'_{2}s'_{2}  \\
   ic'_{1}s'_{1}s_{2}^{'2} & -c_{1}^{'2}s_{2}^{'2} &  ic'_{1}c'_{2}s'_{2} \\
  s'_{1}c'_{2}s'_{2} & ic'_{1}c'_{2}s'_{2}
 & c_{2}^{'2}s_{2}^{'2}  \\ 
\end{array} \right) + m''_{1}\left( \begin{array}{ccc}
  1 & 0 & 0  \\
   0 & 1 & 0 \\
 0 & 0 &  1 \\ 
\end{array} \right)
\end{eqnarray}
and
\begin{eqnarray}
M_{\nu} & = & m_{0}\left( \begin{array}{ccc}
  1 & 0 & 0  \\
   0 & 1 & 0 \\
 0 & 0 &  1 \\ 
\end{array} \right) + 
m'_{0}\left( \begin{array}{ccc}
  s_{1}^{2}s_{2}^{2} & 0 & s_{1}c_{2}s_{2}  \\
  0 & c_{1}^{2}s_{2}^{2} &  0 \\
  s_{1}c_{2}s_{2} & 0
 & c_{2}^{2}s_{2}^{2}  \\ 
\end{array} \right) \nonumber \\
& + & m''_{0}\left( \begin{array}{ccc}
  s_{1}^{'2}s_{2}^{'2} & 0 & s'_{1}c'_{2}s'_{2}  \\
  0 & c_{1}^{'2}s_{2}^{'2} &  0 \\
  s'_{1}c'_{2}s'_{2} & 0
 & c_{2}^{'2}s_{2}^{'2}  \\ 
\end{array} \right) 
\end{eqnarray}
where $s_{1} = \sin \theta_{1} = \sigma_{1}/\sigma_{12}$ and 
$s_{2} = \sin \theta_{2} = \sigma_{12}/\sigma$ 
with $\sigma_{12} = \sqrt{\sigma_{1}^{2} + \sigma_{2}^{2}}$ and 
$\sigma =\sqrt{\sigma_{12}^{2} + \sigma_{3}^{2}}$. Similar definitions 
are for $s'_{1}$ and $s'_{2}$. 

 Note that the two non-diagonal matrices in the mass matrix $M_{e}$ are 
rank one matrices. While it is interesting to observe that when 
the four angles $\theta_{1}$, $\theta_{2}$, $\theta'_{1}$ and $\theta'_{2}$
satisfy the following conditions
\begin{equation} 
\frac{s_{1}}{c_{1}} = \frac{s'_{1}}{c'_{1}}, \qquad \frac{c_{2}}{s_{2}} 
= -\frac{s'_{2}}{c'_{2}}
\end{equation}
which is equivalent to $\sigma'_{1}/\sigma'_{2} 
= \sigma_{1}/\sigma_{2}$, $\sigma'_{12}/\sigma'_{3}=-\sigma_{3}/\sigma_{12}$, the 
two non-diagonal matrices in the mass matrix $M_{e}$ can be simultaneously diagonalized 
by a unitary matrix $U_{e}$ via $M'_{e} = U_{e}^{\dagger} M_{e} U_{e}^{\ast}$. Here
\begin{equation} 
M'_{e}  = \left( \begin{array}{ccc}
  0 & 0 & 0  \\
  0 & m'_{1}& 0  \\
  0 & 0 & m_{1}   \\
\end{array} \right) + m''_{1}U_{e}^{\dagger}U_{e}^{\ast}
\end{equation}
and 
\begin{equation}
U_{e}^{\dagger}=\left( \begin{array}{ccc}
  ic_{1} & -s_{1} & 0  \\
 c_{2}s_{1} & -ic_{1}c_{2} & -s_{2} \\
 s_{1} s_{2} & -ic_{1}s_{2} & c_{2}  \\
\end{array} \right)
\end{equation}
where $U_{e}^{\dagger}U_{e}^{\ast}$ has the following explicit form 
\begin{equation}
U_{e}^{\dagger}U_{e}^{\ast} = \left( \begin{array}{ccc}
  s_{1}^{2}-c_{1}^{2} & 2ic_{1}s_{1}c_{2} & 2ic_{1}s_{1}s_{2}  \\
   2ic_{1}s_{1}c_{2} & c_{2}^{2}(s_{1}^{2}-c_{1}^{2})+s_{2}^{2}
 & c_{2}s_{2}(s_{1}^{2}-c_{1}^{2}) - c_{2}s_{2}  \\
  2ic_{1}s_{1}s_{2} & c_{2}s_{2}(s_{1}^{2}-c_{1}^{2}) - c_{2}s_{2}
 & s_{2}^{2}(s_{1}^{2}-c_{1}^{2})+ c_{2}^{2}  \\ 
\end{array} \right)
\end{equation}
The hierarchical structure of the charged lepton mass implies that 
$m''_{1} << m'_{1} << m_{1}$, it is then not difficult to see that 
the matrix $M'_{e}$ will be further diagonalized by a unitary matrix 
$U'_{e}$ via $D_{e} = U_{e}^{'\dagger} M'_{e} U_{e}^{'\ast} =  
U_{e}^{'\dagger} U_{e}^{\dagger} M_{e} U_{e}^{\ast} U_{e}^{'\ast}$ with 
\begin{equation} 
D_{e}  = \left( \begin{array}{ccc}
  m_{e} & 0 & 0  \\
  0 & m_{\mu}& 0  \\
  0 & 0 & m_{\tau}   \\
\end{array} \right) 
\end{equation}
and 
\begin{equation}
U_{e}^{'\dagger}=\left( \begin{array}{ccc}
  1 + O(m''_{1}/m'_{1}) & iO(m''_{1}/m'_{1}) & iO(m''_{1}/m_{1})  \\
 iO(m''_{1}/m'_{1}) & 1 + O(m''_{1}/m'_{1}) & O(m''_{1}/m_{1}) \\
 iO(m''_{1}/m_{1}) & O(m''_{1}/m_{1}) & 1 +O(m''_{1}/m_{1})  \\
\end{array} \right)
\end{equation}
where $m_{e}= O(m''_{1})$, $m_{\mu}= m'_{1} + O(m''_{1}) $ and 
$m_{\tau}= m_{1} + O(m''_{1})$ define the three charged lepton masses. 
This indicates that the unitary matrix $U'_{e}$ does not significantly 
differ from the unit matrix. Applying the same conditions given in eq.(7), 
the neutrino mass matrix can be rewritten as
\begin{equation}
M_{\nu} = m_{0}\left( \begin{array}{ccc}
  1 + \Delta_{-}s_{1}^{2} & 0 & 2\delta_{-}s_{2}c_{2}  \\
  0 & 1 + \Delta_{-}c_{1}^{2} &  0 \\
  2\delta_{-}s_{2}c_{2} & 0
 & 1 + \Delta_{+}  \\ 
\end{array} \right)
\end{equation}
with 
\begin{equation}
\Delta_{\pm} = \delta_{+} \pm \delta_{-}\cos 2\theta_{2}, 
\qquad \delta_{\pm} = (m'_{0}\pm m''_{0})/2m_{0}
\end{equation}
This neutrino mass matrix can be easily diagonalized by an orthogonal matrix
$O_{\nu}$ via $O_{\nu}^{T}M_{\nu}O_{\nu}$. Explicitly, the matrix $O_{\nu}$
is found to be  
\begin{equation}
O_{\nu}=\left( \begin{array}{ccc}
  c_{\nu} & 0 & s_{\nu}  \\
 0 & 1 & 0 \\
 - s_{\nu} & 0 & c_{\nu}  \\
\end{array} \right)
\end{equation}
with 
\begin{equation}
\tan2\theta_{\nu} = 2\delta_{-}\sin2\theta_{2}/(\Delta_{+} -\Delta_{-}s_{1}^{2})
\end{equation}
When going to the physical mass basis of the neutrinos and charged leptons, 
we then obtain the CKM-type lepton mixing matrix $U_{LEP}$  
appearing in the interactions of the charged weak gauge bosons and leptons, i.e.,
${\cal L}_{W} = \bar{e}_{L}\gamma^{\mu}U_{LEP} \nu_{L} W_{\mu}^{-} + h.c. $. 
Explicitly, we have 
\begin{equation}
U_{LEP} = U_{e}^{'\dagger}U_{e}^{\dagger}O_{\nu} = U_{e}^{'\dagger}\left( \begin{array}{ccc}
  ic_{1}c_{\nu} & -s_{1} & ic_{1}s_{\nu}  \\
 c_{2}s_{1}c_{\nu}+ s_{2}s_{\nu} & -ic_{1}c_{2} & c_{2}s_{1}s_{\nu}-s_{2}c_{\nu} \\
 s_{1} s_{2}c_{\nu}-c_{2}s_{\nu} & -ic_{1}s_{2} & s_{1}s_{2}s_{\nu}+c_{2}c_{\nu}  \\
\end{array} \right)
\end{equation}
The three neutrino masses are found to be
\begin{eqnarray}
m_{\nu_{e}} & = &  m_{0}[ 1 + \frac{1}{2}(\Delta_{+} + \Delta_{-}s_{1}^{2} ) - \frac{1}{2} 
(\Delta_{+} - \Delta_{-}s_{1}^{2})\sqrt{1 + \tan^{2}2\theta_{\nu} } \  ] 
\nonumber \\
m_{\nu_{\mu}} & = &  m_{0}[ 1 + \Delta_{-}c_{1}^{2} ]   \\
m_{\nu_{\tau}} & = &  m_{0}[ 1 + \frac{1}{2}(\Delta_{+} + \Delta_{-}s_{1}^{2} ) + \frac{1}{2} 
(\Delta_{+} - \Delta_{-}s_{1}^{2})\sqrt{1 + \tan^{2}2\theta_{\nu} }\   ] \nonumber 
\end{eqnarray}
for $\tan^{2}2\theta_{\nu} << 1$, masses of the three neutrinos 
are simply given by 
\begin{eqnarray}
m_{\nu_{e}} & \simeq &  m_{0}[ 1 + \Delta_{-}s_{1}^{2}  - 
\frac{1}{4} (\Delta_{+} - \Delta_{-}s_{1}^{2}) \tan^{2}2\theta_{\nu}  \  ] 
\nonumber \\
m_{\nu_{\mu}} & \simeq &  m_{0}[ 1 + \Delta_{-}c_{1}^{2} ]   \\
m_{\nu_{\tau}} & \simeq &  m_{0}[ 1 + \Delta_{+} + 
\frac{1}{4} (\Delta_{+} - \Delta_{-}s_{1}^{2})\tan^{2}2\theta_{\nu} \   ] \nonumber 
\end{eqnarray}  
from which one easily reads off the mass-squared differences 
\begin{eqnarray}
\Delta m_{\mu e}^{2} & = & m_{\nu_{\mu}}^{2} - m_{\nu_{e}}^{2} \simeq  
m_{0}^{2}[\Delta_{-}(c_{1}^{2}-s_{1}^{2}) + \frac{1}{4} (\Delta_{+} - \Delta_{-}s_{1}^{2})
\tan^{2}2\theta_{\nu} ] [2 + \Delta_{-}]  \\
\Delta m_{\tau\mu}^{2} & = & m_{\nu_{\tau}}^{2} - m_{\nu_{\mu}}^{2} \simeq 
m_{0}^{2}[\Delta_{+}- \Delta_{-} c_{1}^{2} + \frac{1}{4} (\Delta_{+} - \Delta_{-}s_{1}^{2})
\tan^{2}2\theta_{\nu}  ] [2 + \Delta_{+} + \Delta_{-} c_{1}^{2} ]  \nonumber 
\end{eqnarray}  

  It is noticed that when $\sin\theta_{\nu} << 1$, we are led to a nearly 2-flavor mixing 
scheme. From the recent atmospheric neutrino data\cite{SUPERK1} which suggested a large neutrino mixing
 between $\nu_{\mu}$ and $\nu_{\tau}$, i.e., the relevant mixing angle satisfies 
$\sin^{2}2\theta > 0.8$, we then obtain the almost same constraint on $\theta_{2}$ when neglecting other 
small mixing angles 
\begin{equation}
\sin^{2} 2\theta_{2} > 0.8
\end{equation} 
Thus the condition $\sin\theta_{\nu} << 1$ is equivalent to $\delta_{-} << \delta_{+}c_{1}^{2}$, 
we have, to a good approximation, the simple relations: $\Delta_{+} \simeq \Delta_{-} \simeq \delta_{+}$ 
and $\tan 2\theta_{\nu}\simeq 2\delta_{-}/\delta_{+}c_{1}^{2}$. 
with this approximation, the neutrino mass-squared differences become more simple
\begin{eqnarray}
\Delta m_{\mu e}^{2} & = & m_{\nu_{\mu}}^{2} - m_{\nu_{e}}^{2} \simeq  
m_{0}^{2}\delta_{+}[c_{1}^{2}-s_{1}^{2} + \delta_{-}^{2}/(\delta_{+}c_{1})^{2} ] 
[2 + \delta_{+} ]
\nonumber \\
\Delta m_{\tau\mu}^{2} & = & m_{\nu_{\tau}}^{2} - m_{\nu_{\mu}}^{2} \simeq 
m_{0}^{2}\delta_{+}[s_{1}^{2} + \delta_{-}^{2}/(\delta_{+}c_{1})^{2} ] 
[2 + \delta_{+}(1 + c_{1}^{2}) ]
\end{eqnarray}  
It has been shown\cite{SUPERK1,BKS,FLMS} that to explain the atmospheric neutrino anomaly, 
the required neutrino mass-squared difference $\Delta m_{\tau\mu}^{2}$ favors the range
\begin{equation}
5\times 10^{-4} eV^{2} < \Delta m_{\tau\mu}^{2} < 6\times 10^{-3} eV^{2}
\end{equation}
To understand the observed deficit of the solar neutrino fluxes in comparison with the 
solar neutrino fluxes computed from the solar standard model\cite{BP}, 
the required neutrino mass-squared difference $\Delta m_{\mu e}^{2}$ falls into the range\cite{BKS}:
\begin{equation}
6\times 10^{-11} eV^{2} < \Delta m_{\mu e}^{2} < 2\times 10^{-5} eV^{2}
\end{equation} 
Here the larger and smaller values of $\Delta m_{\mu e}^{2}$ provide MSW\cite{MSW} 
and just-so\cite{JS} explanations for the solar neutrino puzzle respectively. It is seen that 
the ratio between the two mass-squared differences must satisfy
$\Delta m_{\mu e}^{2}/\Delta m_{\tau\mu}^{2}< 0.04$. With this condition and $\delta_{-}<<\delta_{+}$, 
we then obtain from eq.(22) the following constraint on the mixing angle $\theta_{1}$ 
\begin{equation}
|c_{1}^{2}/s_{1}^{2} - 1| < 0.04
\end{equation}
Note that this constraint is independent of the mass scale $m_{0}$. 
With these constraints, we arrive at the following relations
\begin{equation} 
\frac{m''_{1}}{m'_{1}} \sim \sqrt{\frac{m_{e}}{m_{\mu}}} = 0.07, \qquad  
\frac{m''_{1}}{m_{1}} \sim \frac{\sqrt{m_{e}m_{\mu}}}{m_{\tau}} = 0.004
\end{equation}
Due to the smallness of the mixing angles in $U'_{e}$ and $\theta_{\nu}$, we may
conclude that the neutrino mixing between $\nu_{e}$ and $\nu_{\mu}$ is almost maximal
\begin{equation} 
\sin^{2}2\theta_{1} > 0.998
\end{equation} 
which may leave just-so oscillations as the only viable explanation 
of the solar neutrino data as it can be seen from the analyses in \cite{MY}. 
This requires that $\sigma_{1} \simeq \sigma_{2}$ and
$m'_{0}\simeq m''_{0}$ which may need a fine-tuning if they are not ensured by symmetries.  

   With the above analyses, we may come to the conclusion that with 2-flavor mixing 
and the hierarchical mass-squared differences $\Delta m_{\mu e}^{2}<<\Delta m_{\tau\mu}^{2}$,
the present scheme favors a `bi-maximal' neutrino mixing pattern for 
the explanations of the solar and atmospheric neutrino flux anomalies. 

 It is not difficult to show that the resulting `bi-maximal' neutrino mixing pattern 
allows the three neutrino masses to be nearly degenerate and large enough for hot dark matter
without conflict with the current data on neutrinoless double beta decay. 
This can be seen from the fact that the failure of detecting neutrinoless double beta decay provide 
bounds on an effective electron neutrino mass $<m_{\nu_{e}}> = \sum_{i} 
m_{\nu_{i}}(U_{LEP})_{ei}^{2} < 0.46$ eV\cite{DBD}. To a good approximation, 
when neglecting the small mixing angles in $U'_{e}$,  we obtain
\begin{equation}
 <m_{\nu_{e}}> \simeq m_{0}|s_{1}^{2} - c_{1}^{2}| < 0.46 eV
\end{equation}
Assuming that neutrino masses are large enough to play an essential 
role in the evolution of the large-scale structure of the universe, we may set 
$m_{0} \sim 2$ eV, thus the above constraint will result in the following bound on the 
mixing angle $\theta_{1}$ 
\begin{equation}
|s_{1}^{2}- c_{1}^{2}| < 0.23
\end{equation}
which is weaker than the one given in eq.(25). 

  The smallness of the mass-squared difference $\Delta m_{\mu e}^{2}$ implies that 
$\sin\theta_{\nu} < 0.001$ for $m_{0} \sim 2$ eV. To a good approximation, we may neglect the small 
mixing angle $\theta_{\nu}$ and the small mixing of order $m''_{1}/m_{1}$ in $U'_{e}$. 
With these considerations, the CKM-type lepton mixing matrix is simply given by 
\begin{equation}
U_{LEP} \simeq
\left( \begin{array}{ccc}
  \frac{1}{\sqrt{2}}i & -\frac{1}{\sqrt{2}} & -i\sqrt{\frac{m_{e}}{m_{\mu}}}s_{2}  \\
  \frac{1}{\sqrt{2}}c_{2} & -\frac{1}{\sqrt{2}}c_{2}i & -s_{2} \\
 \frac{1}{\sqrt{2}}s_{2} & -\frac{1}{\sqrt{2}}s_{2}i & c_{2}  \\
\end{array} \right)
\end{equation}
which arrives at the pattern suggested in\cite{FV} 
when neglecting the small mixing angle at the order of magnitude $\sqrt{m_{e}/m_{\mu}}$.
When going back to the weak gauge and charged-lepton mass basis, 
we find that the neutrino mass matrix has the following simple form   
\begin{equation}
M_{\nu} \simeq m_{0}\left( \begin{array}{ccc}
 -\frac{m_{e}}{m_{\mu}}s_{2}^{2}  & ic_{2} & is_{2}  \\
   ic_{2} & s_{2}^{2}
 &  - c_{2}s_{2}  \\
  is_{2} &  - c_{2}s_{2}
 &  c_{2}^{2}  \\ 
\end{array} \right)
\end{equation}

Suggested by the recent atmospheric neutrino data, 
we are motivated to consider two particular interesting cases: Firstly, setting the 
vacuum expectation values to be $\sigma_{3}^{2}
= \sigma_{1}^{2} + \sigma_{2}^{2}$ and $\sigma_{1} = \sigma_{2}$, 
namely, $s_{1}=s_{2}=1/\sqrt{2}$ ($\sin^{2}2\theta_{1} =\sin^{2}2\theta_{2}=1$), 
we then obtain a realistic bi-maximal mixing pattern with 
a maximal CP-violating phase. Explicitly, the neutrino mass and mixing 
matrices read 
\begin{equation}
M_{\nu} \simeq m_{0}\left( \begin{array}{ccc}
  -0.002 & \frac{1}{\sqrt{2}}i & \frac{1}{\sqrt{2}}i  \\
   \frac{1}{\sqrt{2}}i & \frac{1}{2} &  - \frac{1}{2}  \\
  \frac{1}{\sqrt{2}}i &  - \frac{1}{2} &  \frac{1}{2}  \\ 
\end{array} \right)
\end{equation}
and 
\begin{equation}
U_{LEP} \simeq
\left( \begin{array}{ccc}
  \frac{1}{\sqrt{2}}i & -\frac{1}{\sqrt{2}} & -0.05i  \\
  \frac{1}{2} & -\frac{1}{2}i & -\frac{1}{\sqrt{2}} \\
 \frac{1}{2} & -\frac{1}{2}i & \frac{1}{\sqrt{2}}  \\
\end{array} \right)
\end{equation}
when neglecting the small mixing angle at the order of magnitude $\sqrt{m_{e}/m_{\mu}}$, 
we then yield the pattern suggested by  Georgi and Glashow\cite{GG}. 

Secondly, setting the three vacuum expectation values $\sigma_{i}$ (i=1,2,3) 
to be democratic, i.e., $\sigma_{1} = \sigma_{2}=\sigma_{3}$,
hence $s_{1}=1/\sqrt{2}$ and $s_{2}=\sqrt{2/3}$ 
($\sin^{2}2\theta_{1}=1$ and $\sin^{2}2\theta_{2}=0.89$), we 
then arrive at a realistic democratic mixing pattern with a maximal
CP-violating phase. The explicit neutrino mass and mixing matrices 
are given by 
\begin{equation}
M_{\nu} \simeq m_{0}\left( \begin{array}{ccc}
  -0.003 & \frac{1}{\sqrt{3}}i & \frac{2}{\sqrt{6}}i  \\
   \frac{1}{\sqrt{3}}i & \frac{2}{3} &  - \frac{\sqrt{2}}{3}  \\
  \frac{2}{\sqrt{6}}i  &  - \frac{\sqrt{2}}{3} &  \frac{1}{3}  \\ 
\end{array} \right)
\end{equation}
and 
\begin{equation}
U_{LEP} \simeq 
\left( \begin{array}{ccc}
  \frac{1}{\sqrt{2}}i & -\frac{1}{\sqrt{2}} & -0.057i  \\
  \frac{1}{\sqrt{6}} & -\frac{1}{\sqrt{6}}i & - \frac{2}{\sqrt{6}} \\
 \frac{1}{\sqrt{3}} & -\frac{1}{\sqrt{3}}i & \frac{1}{\sqrt{3}}  \\
\end{array} \right)
\end{equation}
when further neglecting the small mixing angle at the order of magnitude 
$\sqrt{m_{e}/m_{\mu}}$, we obtain a similar form provided by Mohapatra\cite{RABI}.   

     From the hierarchical feature in $\Delta m^{2}$, i.e., $\Delta m_{\mu e}^{2}<<
\Delta m_{\tau\mu}^{2}\simeq \Delta m_{\tau e}^{2}$, and the nearly `bi-maximal' mixing pattern, 
formulae for the oscillation probabilities can be greatly simplified to be
\begin{eqnarray}
& & P_{\nu_{e}\rightarrow \nu_{e}}|_{solar} \simeq  1 - 
4|U_{e1}|^{2}|U_{e2}|^{2} \sin^{2}(\frac{\Delta m_{\mu e}^{2}L}{4E}) \nonumber \\
& & P_{\nu_{\mu}\rightarrow \nu_{\mu}}|_{atmospheric} \simeq 1 -4(1-|U_{\mu 3}|^{2})|U_{\mu 3}|^{2}
\sin^{2}(\frac{\Delta m_{\tau\mu}^{2}L}{4E}) \\
& & P_{\nu_{\beta}\rightarrow \nu_{\alpha}} \simeq 4|U_{\beta 3}|^{2}|U_{\alpha 3}|^{2}
\sin^{2}(\frac{\Delta m_{\tau\mu}^{2}L}{4E}) \nonumber 
\end{eqnarray}
and 
\begin{equation}
\frac{P_{\nu_{\mu}\rightarrow \nu_{e}}}{ P_{\nu_{\mu}\rightarrow \nu_{\tau}}}|_{atmospheric}
\simeq \frac{|U_{e 3}|^{2}}{|U_{\tau 3}|^{2}} << 1
\end{equation}
This may present the simplest scheme for reconciling both 
solar and atmospheric neutrino fluxes via oscillations of three neutrinos. 

 On the other hand, the three nearly degenerate neutrino masses can be large enough for hot dark matter.
The relation between the total neutrino mass $m(\nu)$ and the fraction $\Omega_{\nu}$ of 
critical density that neutrinos contribute is \cite{HDM}
\begin{equation}
\frac{\Omega_{\nu}}{\Omega_{m}} = 0.03 \frac{m(\nu)}{1eV} \left(\frac{0.6}{h}\right)^{2} 
\frac{1}{\Omega_{m}} \simeq 0.09 \frac{m_{0}}{1eV} \left(\frac{0.6}{h}\right)^{2} 
\frac{1}{\Omega_{m}}  
\end{equation}
with $h=0.5-0.8$ the expansion rate of the universe (Hubble constant $H_{0}$) in units of 100 km/s/Mpc. 
$\Omega_{m}$ is the fraction of critical density that matter contributes. For $m_{0} \sim 2$ eV and $h=0.6$
the fraction $\Omega_{\nu} \simeq 18 \%$ for $\Omega_{m} = 1$.

 We now come to discuss SO(3) gauge interactions in the present scheme. 
Explicitly, the SO(3) gauge interactions in the mass 
eigenstate of the leptons have the following form 
\begin{equation} {\cal L}_{F} 
= \frac{1}{2} g'_{3} A_{\mu}^{i} \left(\bar{\nu}_{L}t^{i}\gamma^{\mu}\nu_{L} + 
\bar{e}_{L} K_{e}^{i}\gamma^{\mu}e_{L} - \bar{e}_{R} K_{e}^{i 
\ast}\gamma^{\mu}e_{R}\right) 
\end{equation} 
with $K_{e}^{i}=U_{e}^{'\dagger} 
U_{e}^{\dagger}t^{i}U_{e}U'_{e}$. After the SO(3) gauge symmetry is 
spontaneously broken down, the gauge fields $A_{\mu}^{i}$ receive masses by 
`eating' three of the rotational fields. For the SO(3) vacuum structure given 
above, $A_{\mu}^{1}$ and $A_{\mu}^{3}$ are not in the mass eigenstates since 
they mix each other. The mass matrix of the gauge fields $A_{\mu}^{i}$ is found 
to be 
\begin{equation} 
M_{F}^{2} = \frac{1}{4}g^{' 2}_{3}\left( 
\begin{array}{ccc} \sigma_{12}^{2}+\sigma_{12}^{'2}  & 0  & 
-(\sigma_{1}\sigma_{3}+\sigma'_{1}\sigma'_{3}) \\ 0 & 
\sigma_{13}^{2}+\sigma_{13}^{'2} & 0  \\ 
-(\sigma_{1}\sigma_{3}+\sigma'_{1}\sigma'_{3})  &  0 & 
\sigma_{23}^{2}+\sigma_{23}^{'2}  \\ \end{array} \right) 
\end{equation} 
with $\sigma_{ij}^{2} = \sigma_{i}^{2} + \sigma_{j}^{2}$. By using the conditions 
given in eq.(7), the above mass matrix reads 
\begin{equation}
 M_{F}^{2} = m_{F}^{2}\left( \begin{array}{ccc}
 1+ \xi  & 0  & -s_{1}(\frac{c_{2}}{s_{2}} -\frac{s_{2}}{c_{2}}\xi ) \\ 
0 & (s_{1}^{2}+\frac{c_{2}^{2}}{s_{2}^{2}}) 
+(c_{1}^{2}+\frac{s_{2}^{2}}{c_{2}^{2}} ) \xi & 0  \\ 
-s_{1}(\frac{c_{2}}{s_{2}} -\frac{s_{2}}{c_{2}} \xi )   
&  0 & (c_{1}^{2}+\frac{c_{2}^{2}}{s_{2}^{2}}) 
+(s_{1}^{2}+\frac{s_{2}^{2}}{c_{2}^{2}} ) \xi   \\ 
\end{array} \right)  
\end{equation} 
with $ m_{F}^{2}=g^{'2}_{3}\sigma_{12}^{2}/4$ and 
$\xi = \sigma_{12}^{'2}/\sigma_{12}^{2}$. This mass matrix is diagonalized 
by an orthogonal matrix $O_{F}$ via $O_{F}^{T}M_{F}^{2}O_{F}$.  
Denoting the physical gauge fields as $F_{\mu}^{i}$, 
we then have $A_{\mu}^{i} = O_{F}^{ij}F_{\mu}^{j}$. Explicitly, 
\begin{equation} 
\left( \begin{array}{c} A_{\mu}^{1} \\ A_{\mu}^{2} \\ A_{\mu}^{3} \\ \end{array} 
\right)  = \left( \begin{array}{ccc} c_{F} & 0  & -s_{F} \\ 0 & 1 & 0  \\ s_{F}  
&  0 & c_{F}  \\ \end{array} \right) \left( \begin{array}{c} F_{\mu}^{1} \\ 
F_{\mu}^{2} \\ F_{\mu}^{3} \\ \end{array} \right) 
\end{equation} 
with $c_{F}\equiv \cos \theta_{F}$ and $s_{F}\equiv \sin \theta_{F}$. The mixing angle 
$\theta_{F}$ is given by 
\begin{equation}
\tan 2\theta_{F} = \frac{2s_{1}(\frac{c_{2}}{s_{2}} -\frac{s_{2}}{c_{2}}\xi )}{
(\frac{c_{2}^{2}}{s_{2}^{2}}-s_{1}^{2}) +(\frac{s_{2}^{2}}{c_{2}^{2}}-c_{1}^{2} ) \xi }
\end{equation}
Masses of the three physical gauge bosons $F_{\mu}^{i}$ are found to be 
\begin{eqnarray}
m_{F_{1}}^{2} & = &  \frac{m_{F}^{2}}{2} \left( (c_{1}^{2}+\frac{1}{s_{2}^{2}}) 
+(s_{1}^{2}+\frac{1}{c_{2}^{2}} ) \xi - [(\frac{c_{2}^{2}}{s_{2}^{2}}-s_{1}^{2}) 
+(\frac{s_{2}^{2}}{c_{2}^{2}}-c_{1}^{2} ) \xi ] \sqrt{1 + \tan^{2} 2 \theta_{F}} \right) \nonumber \\
m_{F_{2}}^{2} & = &   m_{F}^{2} \left((s_{1}^{2}+\frac{c_{2}^{2}}{s_{2}^{2}}) 
+(c_{1}^{2}+\frac{s_{2}^{2}}{c_{2}^{2}} ) \xi \right) \\
m_{F_{3}}^{2} & = &   \frac{m_{F}^{2}}{2} \left( (c_{1}^{2}+\frac{1}{s_{2}^{2}}) 
+(s_{1}^{2}+\frac{1}{c_{2}^{2}} ) \xi + [(\frac{c_{2}^{2}}{s_{2}^{2}}-s_{1}^{2}) 
+(\frac{s_{2}^{2}}{c_{2}^{2}}-c_{1}^{2} ) \xi ] \sqrt{1 + \tan^{2} 2 \theta_{F}} \right) \nonumber
\end{eqnarray}
For two `bi-maximal' mixing cases considered above, these formulae are simplified to be  
\begin{eqnarray}
\tan 2\theta_{F} & = & \frac{2\sqrt{2}(1-\xi)}{1 + \xi} \nonumber \\
m_{F_{1}}^{2} & = &   \frac{3m_{F}^{2}}{2} \left( \frac{5}{6} - 
\frac{1}{6} \sqrt{1 + \tan^{2} 2 \theta_{F}} \right) (1+ \xi) \nonumber \\
m_{F_{2}}^{2} & = &   \frac{3m_{F}^{2}}{2}( 1 + \xi ) \\
m_{F_{3}}^{2} & = &   \frac{3m_{F}^{2}}{2}\left( \frac{5}{6} + 
\frac{1}{6} \sqrt{1 + \tan^{2} 2 \theta_{F}} \right) (1+ \xi) \nonumber 
\end{eqnarray}
for bi-maximal mixing pattern, i.e., $s_{1} = s_{2} = 1/\sqrt{2}$, and 
\begin{eqnarray}
\tan 2\theta_{F} & = & \frac{2(1-2\xi)}{3\xi} \nonumber \\
m_{F_{1}}^{2} & = &   m_{F}^{2} \left( 1 + \frac{7}{4}\xi 
-\frac{3}{4} \xi\sqrt{1 + \tan^{2} 2 \theta_{F}} \right)  \nonumber \\
m_{F_{2}}^{2} & = &  m_{F}^{2} ( 1 + \frac{5}{2}\xi ) \\
m_{F_{3}}^{2} & = &  m_{F}^{2} \left( 1 + \frac{7}{4}\xi 
+\frac{3}{4} \xi\sqrt{1 + \tan^{2} 2 \theta_{F}} \right)  \nonumber 
\end{eqnarray}
for democratic mixing pattern, i.e., $s_{1} =  1/\sqrt{2}$ and $s_{2} = \sqrt{2/3}$.
In general, the mixing angle $\theta_{F}$ is nonzero and masses of the three gauge bosons 
are splited after spontaneous symmetry breaking. While it is noted that 
for the bi-maximal mixing with $\xi = 1$ and for the democratic mixing  
with $\xi = 1/2$, the mixing angle $\theta_{F}$ will vanish and masses of the two gauge bosons 
$F_{\mu}^{2}=A_{\mu}^{2}$ and $F_{\mu}^{3}=A_{\mu}^{3}$ become degenerate.

   In the physical mass basis of the leptons and gauge bosons, 
the gauge interactions of the leptons are given by the following form 
\begin{equation}
{\cal L}_{F} =  \frac{1}{2}g'_{3} F_{\mu}^{i}\left(\bar{\nu}_{L}t^{j}
O_{F}^{ji}\gamma^{\mu}\nu_{L}
+ \bar{e}_{L} V_{e}^{i}\gamma^{\mu}e_{L} - \bar{e}_{R} V_{e}^{i \ast}
\gamma^{\mu}e_{R}\right) 
\end{equation}
with $V_{e}^{i} = K_{e}^{j}O_{F}^{ji}$. 
To be explicit, we have 
\begin{equation}
K_{e}^{1} = \left( \begin{array}{ccc}
  2c_{1}s_{1} & ic_{2}(s_{1}^{2}-c_{1}^{2}) & is_{2} (s_{1}^{2}-c_{1}^{2}) \\
   -ic_{2}(s_{1}^{2}-c_{1}^{2}) & 2c_{1}s_{1}c_{2}^{2} &  2c_{1}s_{1} c_{2}s_{2}  \\
  -is_{2}(s_{1}^{2}-c_{1}^{2}) &  2c_{1}s_{1} c_{2}s_{2} & 2c_{1}s_{1}s_{2}^{2}  \\ 
\end{array} \right)
\end{equation} 
\begin{equation}
K_{e}^{2} = \left( \begin{array}{ccc}
  0 & c_{1}s_{2} & -c_{1}c_{2}  \\
   c_{1}s_{2} & 0 &  is_{1}  \\
 -c_{1}c_{2} &  -is_{1} & 0 \\ 
\end{array} \right)
\end{equation} 
and 
\begin{equation}
K_{e}^{3} = \left( \begin{array}{ccc}
  0 & is_{1}s_{2} & -is_{1}c_{2}  \\
   -is_{1}s_{2}  & 2c_{1}c_{2}s_{2} &  (s_{2}^{2}-c_{2}^{2})c_{1}  \\
 is_{1}c_{2} &  (s_{2}^{2}-c_{2}^{2})c_{1} & -2c_{1}c_{2}s_{2}  \\ 
\end{array} \right)
\end{equation} 
and 
\begin{eqnarray}
V_{e}^{1} & = & \cos\theta_{F}K_{e}^{1} + \sin\theta_{F}K_{e}^{3}, \nonumber \\
 V_{e}^{2} & = & K_{e}^{2}, \\
V_{e}^{3} & = & -\sin\theta_{F}K_{e}^{1} + \cos\theta_{F}K_{e}^{3} \nonumber 
\end{eqnarray} 
As the mixing matrix $U'_{e}$ does not significantly deviate from the unit matrix,  
the main features in ref.\cite{YLWU} do not change significantly. 
In particular, we will obtain, from the current data on lepton flavor violating process 
$\mu \rightarrow 3e$ with $Br(\mu \rightarrow 3e) < 1\times 10^{-12}$\cite{LFV},
 a similar constraint on the SO(3) symmetry breaking scale 
\begin{equation}
\sigma_{12} > 10^{3} v 
\frac{m_{F}\sqrt{m_{F_{3}}^{2} -m_{F_{1}}^{2}}}{m_{F_{1}}m_{F_{3}}} s_{1}\sqrt{c_{1}s_{2}}
\end{equation}
with $v=246$ GeV the weak symmetry breaking scale. Specifically, we have 
\begin{equation}
\sigma_{12} > 10^{3} \frac{v}{2\sqrt{3}}
 \left(\frac{\tan2\theta_{F} (1 + \frac{1}{2\sqrt{2}}\tan2\theta_{F}) }{1-
\frac{1}{24}\tan^{2}2\theta_{F} } \right)^{1/2}
\end{equation}
for bi-maximal mixing case, and 
\begin{equation}
\sigma_{12} > 10^{3} \frac{v}{3}
 \left(\frac{\tan2\theta_{F} (1 + \frac{3}{4}\tan2\theta_{F}) }{\sqrt{3}(1 + 
\frac{5}{6}\tan2\theta_{F} + \frac{1}{8}\tan^{2}2\theta_{F} } \right)^{1/2}
\end{equation}
for democratic mixing case. When the mixing angle $\theta_{F}$ is at the same order of 
magnitude as the weak mixing angle $\theta_{W}$, by setting $\tan 2\theta_{F}\simeq 3/2$, we then obtain
\begin{equation}
\sigma_{1} \simeq \sigma_{2}\simeq \sigma_{3}/\sqrt{2} > 0.33\times 10^{3}\ v 
\simeq 81\ TeV 
\end{equation}
for bi-maximal mixing case, and 
\begin{equation}
\sigma_{1} \simeq \sigma_{2}\simeq \sigma_{3} > 0.2\times 10^{3}\ v 
\simeq 49\ TeV 
\end{equation}
for democratic mixing case. Suppose that the SO(3) gauge coupling 
constant $g'_{3}$ is at the same order of magnitude as the electroweak coupling constant $g$, 
masses of the three SO(3) gauge bosons are bounded for $\theta_{F} \sim \theta_{W}$ to be 
\begin{equation}
m_{F_{1}} > 38\ TeV, \qquad m_{F_{2}} > 53\ TeV , \qquad m_{F_{3}} > 57\ TeV
\end{equation}
for bi-maximal mixing case, and 
\begin{equation}
m_{F_{1}} > 24\ TeV, \qquad m_{F_{2}} > 29\ TeV , \qquad m_{F_{3}} > 32\ TeV
\end{equation}
for democratic mixing case. When the mixing angle becomes very small $\theta_{F} << 1$,  
the constraint on the SO(3) symmetry breaking scale is approximately given by 
\begin{equation}
\sigma_{1} \simeq \sigma_{2}\sim \sigma_{3} > 45\sqrt{\tan 2\theta_{F}}\ TeV 
\end{equation}
Once the mixing angle $\theta_{F} $ is extremely small at the order of magnitude  
$\sin\theta_{F} \sim 10^{-4}$, the SO(3) symmetry breaking scale can be below 1 TeV 
and the SO(3) gauge boson masses may reach to the order of magnitude 300 GeV.

   In summary, we have investigated a realistic scheme for lepton masses and mixings 
within the framework of the gauged SO(3) lepton flavor symmetry discussed recently 
in ref.\cite{YLWU}. A nearly `bi-maximal' neutrino mixing pattern 
with maximal CP-violating phase has been derived to explain the  
solar and atmospheric neutrino data reported recently by the Super-Kamiokande 
experiment when LSND results\cite{LSND} are not considered. 
This is because including the LSND results, it likely needs to introduce a 
sterile neutrino\cite{STERILE}. We has also shown that due to
the intriguing feature of the vacuum structure of spontaneous SO(3) gauge 
symmetry breaking, the three Majorana neutrino masses in the scheme are allowed to 
be nearly degenerate and large enough for a hot dark matter candidate.
Though neutrinoless double beta decay may become unobservable small, 
the scheme still allows rich interesting phenomena on lepton flavor violations via the 
SO(3) gauge interactions.

 {\bf Ackowledgments}: This work was supported in part by the NSF of China under the 
grant No. 19625514. The author would like to thank Alexei Smirnov for bringing his attention 
on the most recent paper by R. Barbieri, et al. \cite{BHKR}.


\end{document}